# $\gamma$-ray detection from occasional flares in T Tauri stars of NGC 2071 – I. Observational connection

A. Filócomo,[1,2]★ J. F. Albacete-Colombo,[1] E. Mestre,[3,4] L. J. Pellizza[5] and J. A. Combi[2,3,6]

[1]*Departamento de Investigación en Ciencias Exactas e Ingeniería, UNRN – Sede Atlántica, Don Bosco y Leloir s/n, 8500 Viedma, Argentina*
[2]*Facultad de Ciencias Astronómicas y Geofísicas, UNLP, Paseo del Bosque s/n, 1900 La Plata, Argentina*
[3]*Departamento de Física (EPSJ), Universidad de Jaén, Paraje de las Lagunillas s/n, E-23009 Jaén, Spain*
[4]*Institute of Space Sciences (ICE, CSIC), Campus UAB, Carrer de Can Magrans s/n, E-08193 Barcelona, Spain*
[5]*Instituto de Astronomía y Física del Espacio (CONICET-UBA), C.C. 67, Suc. 28, C1428ZAA Buenos Aires, Argentina*
[6]*Instituto Argentino de Radioastronomía (CCT La Plata, CONICET), C.C. 5, Villa Elisa, 1894 Buenos Aires, Argentina*

## ABSTRACT

NGC 2071 is a star-forming region that overlaps with three $\gamma$-ray sources detected by the *Fermi Space Telescope*. We propose that strong flare activity in T Tauri stars could produce $\gamma$-ray emission in a way that makes them a counterpart to some unidentified sources detected by the Large Area Telescope aboard the *Fermi* satellite. We have performed a spectral and temporal analysis for two *Fermi* data sets: the first 2 yr and the entire 14 yr of observations. We have found that the $\gamma$-ray source is detectable at 3.2$\sigma$ above the background at energies above 100 GeV during the first 2 yr of observation. The analysis of the expected frequency of the highest energy flares occurring in T Tauri stars is consistent with our estimate. In addition, we have determined the minimum energy of the flare that would produce $\gamma$-ray emission, which is $\sim 5 \times 10^{37}$ erg. This agreement becomes a hard observational constraint supporting previous hypotheses about rare flares as the origin of unidentified $\gamma$-ray sources in star-forming regions.

## 1 INTRODUCTION

The *Fermi Gamma-ray Space Telescope*, with the Large Area Telescope (LAT) onboard, has been studying the Universe in the energy range from 20 MeV to more than 300 GeV since 2008, providing high-energy $\gamma$-ray surveys with unprecedented sensitivity. Based on 14 yr of observations, four catalogues of point sources have been published to date. The first one contains nearly 1500 sources, while the latest shows more than 6600 sources, all with a significance of more than 4$\sigma$ above the background (Abdollahi et al. 2022). Most *Fermi*-LAT sources have been identified at other frequency bands, from radio to X-ray energies, with different types of $\gamma$-ray emitters (such as pulsars and pulsar wind nebulae, X-ray binaries, supernova remnants, radio galaxies, or active galactic nuclei). However, many of them have no clear counterpart at lower frequencies, partly due to the low spatial resolution achieved at high $\gamma$-ray energies. It is noticeable that at least 27 per cent of the sources are classified as unidentified at other frequencies in all catalogues. Understanding the nature of these unknown sources is one of the most important problems in $\gamma$-ray astronomy today.

We focused our study on the origin of unidentified $\gamma$-ray sources in several nearby star-forming regions (SFRs) and found a significant positional correlation with the dark molecular cloud Lynds 1630. This region is located in the northern part of the Orion B molecular cloud and was recognized by Strom et al. (1975) as the site of recent star formation. These authors have also identified many embedded objects, including early-type main-sequence stars, several H$\alpha$ emission line stars, and Herbig–Haro objects. Lynds 1630 includes the SFR NGC 2071, located at a distance of 390 pc (Anthony-Twarog 1982), where the unidentified $\gamma$-ray emission detected by *Fermi*-LAT is centred. This SFR and the cluster NGC 2071-IR, located 4 arcmin north of NGC 2071, have been extensively analysed in infrared (IR; e.g. Lada et al. 1991; Walther et al. 1993; Gibb 2008; Walther & Geballe 2019). The cluster contains Herbig–Haro objects (Zhao et al. 1999) and bipolar molecular outflows (e.g. Eislöffel 2000), indicating clear signs of recent active star formation.

X-ray emission is widespread in low-mass young stellar objects (YSOs) and is orders of magnitude greater than in most main-sequence stars. T Tauri stars typically exhibit rapid, high-amplitude variability and hard X-ray spectra associated with violent magnetic reconnection events (Feigelson & Montmerle 1999). The most notable types of variability are rotational modulations and flare-like activity, the latter exhibiting short-rise phases with up to a 100-fold increase in flux from pre-flare levels and slower decay phases over many hours (see e.g. Favata et al. 2005). Frequent flares indicate magnetic confinement and, in most cases, heating of the plasma by magnetic reconnection. These conditions may lead to the emergence of a significant population of non-thermal relativistic particles that could emit $\gamma$-ray radiation.

The solar flare model (Reale 2003) can be used as a first approximation to explain these young stellar flares. Although the environment and properties of T Tauri stars change greatly during their lifetime (from class I to class III), no change in X-ray properties has been observed between different evolutionary stages (see e.g. Favata et al. 2005; Waterfall et al. 2019). Models suggest the existence of very large flare structures in these stars. In several cases, the magnetic structures containing the plasma are thought to be much larger than the stars themselves, which has never been observed in more evolved stars (Favata et al. 2005; Waterfall et al. 2019; Getman, Feigelson & Garmire 2021).

★ E-mail: afilocomo@unrn.edu.ar

The most complete survey for X-ray variability in young stars comes from Getman & Feigelson (2021), who analysed a data set of flares from more than 24 000 young stellar objects in 40 SFRs. They called *megaflares* the most energetic flares in the X-ray band, with $10^{36.2} < E_X < 10^{38}$ erg, where the upper limit stands because no flare was observed in their data set with $\log(E_X/\mathrm{erg}) > 38$. Megaflares are characterized by a mean loop height $L \sim 0.8 R_\star$, where $R_\star$ is the stellar radius, a typical loop cross-sectional area of $10^{21}$ cm$^2$, an equipartition magnetic field in the flare $B_{eqp} \sim 200$ G, and a photospheric magnetic field $B_{ph} \sim 1$ kG. These properties make them ideal candidates for accelerating particles to relativistic energies and thus places where γ-rays could be emitted. The frequency of X-ray flares is a function of their energy and the mass of the star. For a low-mass ($M < 1$ M$_\odot$) stellar population, the frequency $N(E_X)$ of megaflares with energy greater than $E_X$ per star per year is

$$\log\left[\frac{N(E > E_X)}{(\mathrm{star\ yr})^{-1}}\right] = \kappa - \beta \log\left(\frac{E_X}{\mathrm{erg}}\right), \quad (1)$$

where $\kappa = 46.94$, $\beta = 1.31$, and the uncertainty of the estimate is $\pm 0.2$ dex, according to Getman & Feigelson (2021).

The first serious hypothesis that at least part of the γ-ray emission observed by the *Fermi* satellite in some nearby SFRs may have its origin in flares from T Tauri stars comes from del Valle et al. (2011). These authors found a positional match between an unidentified *Fermi* source and four T Tauri stars and presented a simplified model to explain the radiation produced by non-thermal relativistic particles in the vicinity of these stars. They modelled magnetic fields in the stellar disc system with parameters consistent with those of megaflares, opening a new interpretation for the origin of the observed γ-ray emission in SFRs. The parameters measured by Getman et al. (2021) for the X-ray flares in the stars of the Orion Nebula cluster strongly confirm this idea. However, the flare activity in young stars decreases with the total flare energy (Albacete Colombo et al. 2007), making it difficult to detect a large number of events needed to test the proposed scenario.

In this paper, we present for the first time a convincing argument, based not only on positional coincidence but also on γ-ray spectra and variability, for a population of T Tauri stars undergoing megaflares powering a *Fermi* source. In Section 2, we present the reduction of the *Fermi*-LAT γ-ray data, while in Section 3, we discuss the possible astrophysical origin of the observed emission. Finally, in Section 4, we summarize our results and draw our conclusions.

## 2 γ-RAY ANALYSIS

*Fermi*-LAT (Atwood et al. 2009) operates primarily in an all-sky scanning survey mode, providing 30 min of live time on each point in the sky every two orbits (approximately 3 h). γ-ray emission at GeV energies from NGC 2071 was reported in the first three *Fermi*-LAT source catalogues (Abdo et al. 2010; Nolan et al. 2012; Acero et al. 2015), corresponding to one (1FGL), two (2FGL), and four (3FGL) years of observations, respectively. Fig. 1 shows the *Fermi* error ellipses of those detections. None of these sources have been identified with an observational counterpart at another frequency in the electromagnetic spectrum. The error ellipses of the observed emission in the first two catalogues are smaller and better centred on NGC 2071. Also, no γ-ray source associated with the region of NGC 2071 is found in the *Fermi*-LAT Fourth Source Catalogue (4FGL; Abdollahi et al. 2020), Data Release 2 (4FGL-DR2; Ballet et al. 2020), or Data Release 3 (4FGL-DR3; Abdollahi et al. 2022). The 4FGL catalogue, however, does not contain γ-ray transients or sources only detectable over a short period (since their signal results diluted over the 8 yr of data).

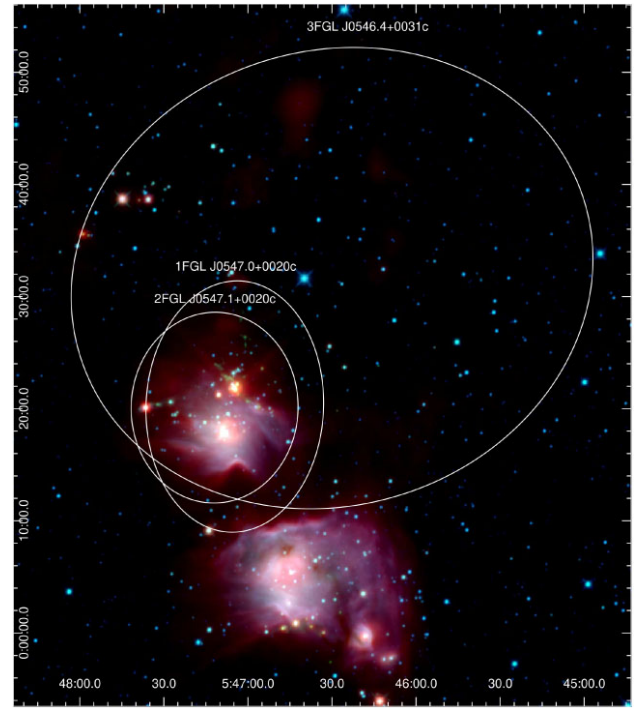

**Figure 1.** NGC 2071 (the nebula close to the centre of the image) obtained with the *Wide-field Infrared Survey Explorer* (*WISE*) using the 22 μm (red), 4.6 μm (green), and 3.4 μm (blue) bands. In white, we show the 3σ significance *Fermi* error ellipses that positionally coincide with NGC 2071. 1FGL, 2FGL, and 3FGL are the first, second, and third *Fermi* catalogue, respectively.

We performed a dedicated analysis of the *Fermi*-LAT data of NGC 2071 by using 2 yr of observations taken between 2008 August 4 and 2010 July 31, corresponding to the observation period of the 2FGL catalogue, since the flux of the 2FGL source reported in the catalogue is the most intense of the three sources. We selected the dubbed P8R3 SOURCE class events (Bruel et al. 2018), in the energy range between 100 MeV and 300 GeV, in a region of interest (ROI) of 20° radius centred on the 2FGL source's position (i.e. 2FGL J0547.1+0020c; RA = 86°.799507, Dec. = 0°.33487). The analysis consists of ROI model optimization, localization and spectrum fitting, and variability study of the source. We used the FERMIPY (version 1.2) PYTHON package that facilitates analysis of LAT data with the *Fermi* Science Tools (Wood et al. 2017). We binned the data into squared spatial bins with 0°.1 size and eight bins per energy decade. Because the Earth's limb is an intense source of background γ-rays, we excluded the detected emission with a zenith angle larger than 90°. The P8R3_SOURCE_V3 instrument response functions allow us to describe the point-like sources located within a 20° radius from the source position. We employed the same source templates[1] used to compile the 4FGL-DR3 catalogue, including those for the Galactic and isotropic diffuse emission. We modelled the ROI according to the 4FGL-DR3, but no source matches NGC 2071. Hence, we included a putative point-like source at the position of 2FGL J0547.1+0020c. A typical power-law model was fitted to the source spectrum given by

$$\frac{dN}{dE} = N_0 \left(\frac{E}{E_0}\right)^\gamma, \quad (2)$$

with the following initial parameters: $N_0 = 10^{-15}$, $\gamma = -0.5$, and a fixed value for the reference energy $E_0 = 2 \times 10^4$ MeV. We applied the energy dispersion correction except for the isotropic diffuse emission.

---
[1] https://fermi.gsfc.nasa.gov/ssc/data/access/lat/BackgroundModels.html

We carried out the analysis by performing a likelihood fit of all free parameters of the ROI model. Then, for each source, we computed a maximum likelihood test statistic (TS) that compares the maximum likelihood of the baseline ROI model without the source, and the same but with the source included. The TS is distributed like $\chi^2$ and a Gaussian approximation can be used to evaluate it, approximating $\sqrt{TS}$ as the Gaussian width. We left free to vary all spectral parameters of isotropic and Galactic diffuse emission models, as well as those for sources with TS > 10, and the normalization of all sources within 3° of the ROI centre. The spectral fit was performed over the energy range from 100 MeV to 300 GeV. As a result, a source with TS = 10.25 was detected, corresponding to a significance of $3.2\sigma$, centred on (RA, Dec.) = (86°.518 ± 0°.161, 0°.013 ± 0°.0161). The spectral best fit for the source emission is adequately described (with only statistical errors) with a power-law energy distribution given by

$$\frac{dN}{dE} = (1.41 \pm 0.03) \times 10^{-16} \left(\frac{E\,[\mathrm{MeV}]}{2 \times 10^4}\right)^{-0.43 \pm 0.01} \mathrm{cm}^{-2}\,\mathrm{s}^{-1}\,\mathrm{MeV}^{-1}. \quad (3)$$

It would be interesting to compare our results with those of a hypothetical active galactic nucleus (AGN), an extragalactic source emitting by chance alignment with our stellar $\gamma$-ray source. AGNs present typical spectral indices above 0.5 at energies between 0.1 and 100 GeV (Dermer & Giebels 2016), while in equation (3) we found a spectral index of 0.4 for our analysis. Thus, except for rare alignments, the intrinsic nature of the observed emission cannot be explained by extragalactic background sources.

According to the obtained model, we computed the spectral energy distribution (SED) shown in Fig. 2 for the source. This figure shows that the source model and its detection are only significant at energies above 100 GeV, near the limit of spectral coverage of the *Fermi* 2FGL catalogue (300 GeV). For $\gamma$-ray energies lower than 100 GeV we obtained upper limits (see Fig. 2). Additionally, we computed the integrated energy flux for this period finding $F_{\mathrm{2FGL}}$ of $1.86 \pm 0.08 \times 10^{-11}$ erg cm$^{-2}$ s$^{-1}$ for energies between 100 and 300 GeV, resulting in an integrated $\gamma$-ray luminosity $L_{100-300\,\mathrm{GeV}}$ of $2.69 \pm 0.11 \times 10^{31}$ erg s$^{-1}$.

## 3 TIME VARIABILITY ANALYSIS

In this paper, we are primarily interested in the temporal behaviour of the source. Unfortunately, the low photon statistics in $\gamma$-rays avoids usual statistical tests to give a reliable estimation of variability. Therefore, we obtained a light curve directly by dividing the observation time interval in bins $\Delta t_i$ and computing the signal under the assumption of constant flux from the source. The mean significance of the detection expectedly increases with time as $\sqrt{TS} \propto \sqrt{\Delta t_i}$. In the best-case scenario, we obtained TS = 10.25 for 2 yr of observations, so we adopted a bin size of 1.75 yr. Since the observation time of 2 yr is too short to run a light curve with a bin size of 1.75 yr, we decided to perform another analysis using all the available *Fermi*-LAT data of NGC 2071, from 2008 August 4 to 2022 October 26. To do this, we used the same background models and the sources listed in 4FGL-DR3 to describe the point-like sources. For the source of interest, we added a point-like source with a typical power-law spectrum. As the $\gamma$-ray emission of the source decreases over the 14 yr of observations, we adopted different tests for the power-law spectral model. We inferred that the best set of initial parameters is $N_0 = 10^{-19}$, $\gamma = -0.5$, and a fixed value for the reference energy $E_0 = 2 \times 10^4$ MeV. We found a low significance for the source over the background ($2.26\sigma$). Then we obtained the light curve adopting a total of eight bins, each with a size of 1.75 yr, which is shown in Fig. 3. The source was detected during the first bin with a significance of $3\sigma$, and remained undetected after that, which strongly suggests that the $\gamma$-ray emission was produced in a transient event, as it is expected from megaflare episodes in T Tauri stars.

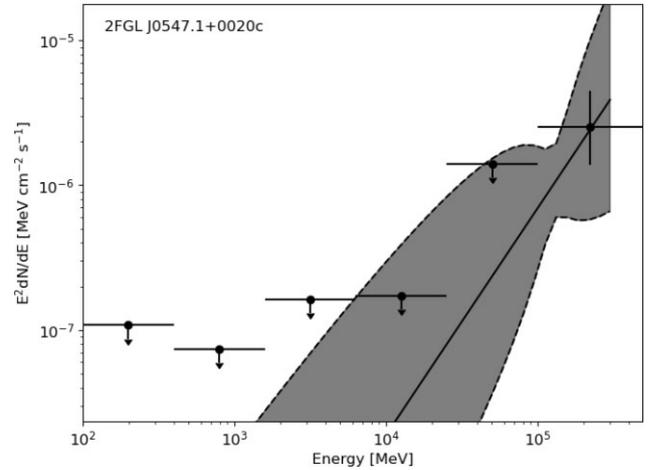

**Figure 2.** Six bin computed SED for the source for 2 yr of observations. The integrated flux in energy is shown by dots and horizontal bars. The shadowed grey area shows the $1\sigma$ confidence region.

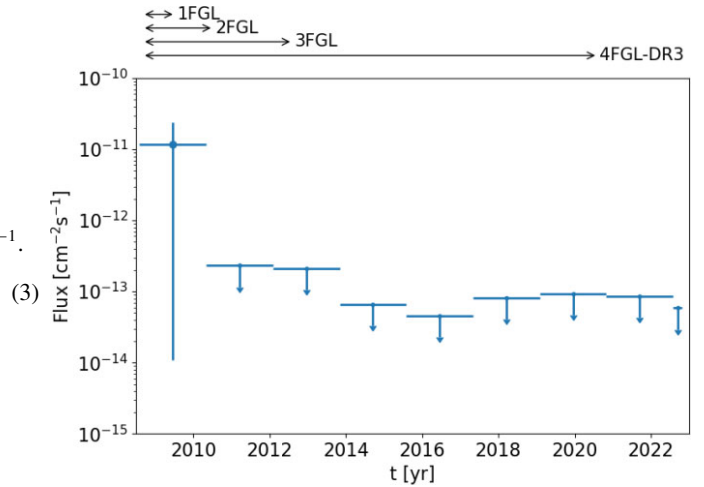

**Figure 3.** Light curve with a 1.75-yr size bin for the source detected within 14.23 yr of *Fermi*-LAT observations. The arrows above indicate the observation period for each *Fermi* catalogue.

The fact that the source is detected in only one of the eight bins of the light curve allows us to draw some conclusions on the frequency of flares of this object, and therefore on their energy. Several stars in the neighbourhood were identified as T Tauri stars using IR analysis (Flaherty & Muzerolle 2008; Gibb 2008; Skinner et al. 2009; Spezzi et al. 2015). At least $N_\star = 58$ stars classified as T Tauri at any stage of evolution are found within the second *Fermi* catalogue's error ellipse. This is used to make a quantitative estimate. Using equation (1), the frequency $\nu(E_X)$ of flares with energies above some threshold value $E_X$ in the area enclosed by the error ellipse is then

$$\log\left[\frac{\nu(E_X)}{\text{yr}^{-1}}\right] = \log\left[N_\star \frac{N(E > E_X)}{\text{yr}^{-1}}\right] = \kappa' - \beta \log\left(\frac{E_X}{\text{erg}}\right), \quad (4)$$

where $\kappa' = \kappa + \log N_\star = 48.70$. Since there are no deep enough X-ray observations to constrain the minimum energy of the flares, we estimate $\nu$ (and hence $E_X$) using the light-curve data.

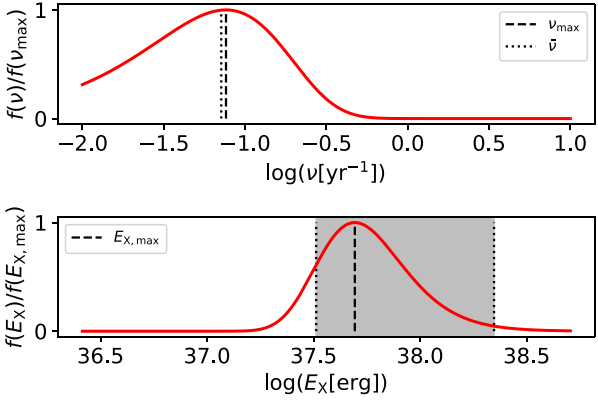

**Figure 4.** Upper panel: posterior PDF of the frequency of flares $\nu$ (solid red line). The value that maximizes the PDF (black dashed line) and the mean frequency of flares in the total observation time (black dotted line), assuming that a single one is observed in $\Delta t_1$, are also shown. Lower panel: posterior PDF of the minimum energy of flares $E_X$ (solid red line), together with the value that maximizes it (dashed black line) and the 90 per cent credible region for $E_X$ (grey shaded region between black dotted lines).

Bayesian statistics allows us to define a probability density function (PDF) $f(\nu|D)$, in such way that $f(\nu|D)\,d\nu$ represents the degree of belief in the proposition that the true value of the frequency lies in $[\nu, \nu + d\nu]$, given the data $D$. The value $\nu_{max}$ that maximizes $f$ is then an estimate of the true frequency. Bayes theorem allows to compute $f(\nu|D)$ as

$$f(\nu|D) = \frac{\mathcal{L}(D|\nu)g(\nu)}{\int_0^\infty \mathcal{L}(D|\nu)g(\nu)\,d\nu}, \quad (5)$$

where the likelihood $\mathcal{L}(D|\nu)$ is the probability of measuring the data $D$ given $\nu$, and $g(\nu)$ is the prior PDF of $\nu$, representing the degree of belief in $\nu$ before taking the data.

Counting the number of flares $k$ in a time interval $\Delta t$ is a Poisson process. Considering the time period of the first eight bins of the light curve, our data $D$ can be described as two independent observations: in the first of them $\Delta t_1 = 1.75$ yr and $k_1 > 0$ (unknown), whereas in the second one $\Delta t_2 = 12.25$ yr and $k_2 = 0$. The total time span of the observations is $\Delta t = 14$ yr. Therefore,

$$\mathcal{L}(D|\nu) = (1 - e^{-\nu \Delta t_1})\,e^{-\nu \Delta t_2}. \quad (6)$$

We adopt a uniform prior PDF $g(\nu) = 1/\Delta \nu$, where $\Delta \nu$ is the width of the considered frequency range. This choice corresponds to the assumption of a complete lack of knowledge on $\nu$ before taking the data. The value of $\Delta \nu$ is irrelevant, as in this case $g(\nu)$ cancels out in equation (5). Moreover, as the denominator in the right-hand side of this equation does not depend on $\nu$, it does not affect the results of the maximization of $f(\nu|D)$.

In Fig. 4, we show the behaviour of $f(\nu|D)$ and $f(E_X|D) = \beta E_X^{-1} \nu f(\nu|D)$, which is the posterior probability density of $E_X$. Using the values adopted for the parameters, we find $\nu_{max} = 0.076$ yr$^{-1}$ (one event every 13.2 yr). The estimated value of the minimum energy of the flares is $E_{X,\,max} = 4.89 \times 10^{37}$ erg, which lies close to the upper limit of the energy range of megaflares observed by Getman & Feigelson (2021). Moreover, the 90 per cent credible region for $E_X$ is the interval $[0.33, 2.22] \times 10^{38}$ erg, which is completely within energy range of megaflares. These results make a strong case for the occurrence of megaflares (as opposed to lower energy eruptions) in the source. The frequency that maximizes the posterior PDF is very similar to that obtained from the mean number of events in the total time interval, $\bar{\nu} = (\Delta t)^{-1} = 0.071$ yr$^{-1}$, under the assumption that there is only one flare during $\Delta t_1$.

To test this last assumption, we compute the betting odds $q$ of observing more than one flare in $\Delta t_1$ against observing exactly one flare, both given the estimated most probable frequency and given that there is at least one flare observed in this bin:

$$q = \frac{P(k > 1|k > 0, \nu_{max})}{P(k = 1|k > 0, \nu_{max})} = \frac{1 - e^{-\nu_{max}\Delta t_1}(1 + \nu_{max}\Delta t_1)}{\nu_{max}\Delta t_1\,e^{-\nu_{max}\Delta t_1}}. \quad (7)$$

From our data we obtain $q = 0.069$, implying a 93.5 per cent probability that the source emits one flare against 6.5 per cent that it emits more than one, in both cases given the most probable frequency, and given that some emission is detected.

It is important to note that our estimates are conservative regarding the number of T Tauri stars present in the source. The adopted value $N_\star = 58$ is a lower limit, as there may be more unidentified stars of this type. In that case, the values of $\kappa'$ in equation (4) would be higher, and the frequency of flares for any $E_X$ would be greater than in the adopted case. However, the frequency estimated from the data would remain the same (as it does not depend on $N_\star$) and therefore it would correspond to a higher value of $E_X$. In other words, the larger the number of T Tauri stars in the region, the rarer and more energetic their flares should be to explain the observations.

## 4 SUMMARY AND CONCLUSIONS

We have found a fading $\gamma$-ray source that positionally coincides with NGC 2071. Three $\gamma$-ray sources detected, respectively, in the first three *Fermi* catalogues, which correspond to the first 4 yr of observations, overlap with this region. The error ellipses in the first two catalogues (1FGL and 2FGL) are smaller and better centred on NGC 2071, whereas in the third catalogue (3FGL) a source with a much larger error in its position is detected, suggesting that the emission is not as intense. Furthermore, in the following years of observations, *Fermi* did not detect any sources in the region. This suggests that the detected $\gamma$-ray emission corresponds to a source that is not constant over time.

We used the most recent isotropic spectral and Galactic interstellar emission models to reanalyse *Fermi* data from the first 2 yr of observations. We detected a source with a significance of $3.2\sigma$ centred in (RA, Dec.) = ($86°.518 \pm 0°.0161$, $0°.013 \pm 0°.061$). Despite the sources in the *Fermi* catalogues are detected with a significance greater than $4\sigma$, it makes sense that we detect the source with a lesser significance since we employed different (updated) background models. According to the SED computed, the source emission is significant at $E \gtrsim 100$ GeV, far from the energy regime (of few GeV) where LAT achieves the best sensitivity. Furthermore, after analysing the data from the whole period of 14 yr of observations available, we could confirm from the light curve that the source only emitted during the first 2 yr of observations.

NGC 2071 is a nearby SFR with a population of low-mass protostars (Anthony-Twarog 1982). According to del Valle et al. (2011), $\gamma$-ray radiation could only come from relativistic particles accelerated in very energetic flares produced by magnetic reconnection in T Tauri stars. Megaflares are very rare events, as Getman & Feigelson (2021) have shown that the frequency of flares decreases with their energy. Under the above scenario, our quantitative Bayesian analysis of the temporal behaviour of the source combined with the energy–frequency relation of Getman & Feigelson (2021) strongly suggests that megaflares are the driver of the $\gamma$-ray emission observed at NGC 2071. Our best estimate for the frequency is one event every 13.2 yr, which is consistent with flare energies well above $10^{37}$ erg and up to several times $10^{38}$ erg, with a very high (90 per cent) confidence level.

This is the first time that mere observational results have shown that the highest energy flares in T Tauri stars may be responsible for some of the unidentified $\gamma$-ray sources detected by the *Fermi* satellite. Although this is at present the only consistent scenario explaining observations, it requires further verification to reinforce the link between T Tauri stars and transient $\gamma$-ray sources. A step forward could be made by devising coordinated multifrequency campaigns to simultaneously detect megaflares (in X-rays) and $\gamma$-ray emission. The detection of radiation coming from non-thermal particles at other wavelengths (e.g. radio) would also help prove this scenario. So, given our results, its proximity, and its position in the sky, NGC 2071 would be an excellent target for such a campaign. However, we could gain much insight by also searching the *Fermi* catalogues for other transient sources whose position is consistent with other SFRs, and performing this analysis on those sources that show clear signs of transient occurrence. In this way, we could expand the sample and provide further evidence for the T Tauri scenario for $\gamma$-ray emission from SFRs.

Remarkably, the energy range in which the LAT source was detected (hundreds of GeV) and the calculated frequency of megaflares per star per year make SFRs such as NGC 2071 interesting targets for future observations with the Cherenkov Telescope Array (CTA, currently under construction), especially for the future Large Size Telescopes (LSTs) and Medium Size Telescopes (MSTs) subarrays and the currently operating LST-1 (Mazin 2021). In addition, deep observations of SFRs are planned as part of the CTA key programs (Cherenkov Telescope Array Consortium et al. 2019).

Finally, this is the first in a series of observational and theoretical papers. The second paper will present a theoretical model for the $\gamma$-ray SED of megaflares in T Tauri stars, which will be compared with observations from the *Fermi Space Telescope*.

## ACKNOWLEDGEMENTS

JFA-C, as a CONICET researcher, and AF, as a CONICET doctoral fellow, acknowledge the support of CONICET and PI-40-C-866 (UNRN). JAC is member of CONICET. EM acknowledges support by grant P18-FR-1580 from the Consejería de Economía y Conocimiento de la Junta de Andalucía under the Programa Operativo FEDER 2014–2020. EM acknowledges support from the I + D + i subproject 'Tecnologías avanzadas para la exploración del universo' from the 'Plan de Recuperación, Transformación y Resiliencia' of the Spanish government (EU Next Generation PRTR-C17.I1 and Generalitat de Catalunya). LJP acknowledges support from Argentine ANPCyT through grant PICT 2020-0582. JAC is a María Zambrano researcher fellow funded by the European Union – NextGenerationEU – (UJAR02MZ). This research was partially funded by CONICET grants PIP 112-201207-00226 and PIP 112-201701-00604. JAC was supported by PIP 0113 (CONICET), PICT-2017-2865 (ANPCyT), and PID2019-105510GB-C32/AEI/10.13039/501100011033 from the Agencia Estatal de Investigación of the Spanish Ministerio de Ciencia, Innovación y Universidades, and by Consejería de Economía, Innovación, Ciencia y Empleo, Junta de Andalucía as research group FQM-322, as well as FEDER funds.

## DATA AVAILABILITY

All *Fermi*-LAT data are public and available at https://fermi.gsfc.nasa.gov/ssc/data/access/lat/.

## REFERENCES

Abdo A. A. et al., 2010, ApJS, 188, 405
Abdollahi S. et al., 2020, ApJS, 247, 33
Abdollahi S. et al., 2022, ApJS, 260, 53
Acero F. et al., 2015, ApJS, 218, 23
Albacete Colombo J. F., Caramazza M., Flaccomio E., Micela G., Sciortino S., 2007, A&A, 474, 495
Anthony-Twarog B. J., 1982, AJ, 87, 1213
Atwood W. B. et al., 2009, ApJ, 697, 1071
Ballet J., Burnett T. H., Digel S. W., Lott B., 2020, preprint (arXiv:2005.11208)
Bruel P., Burnett T. H., Digel S. W., Johannesson G., Omodei N., Wood M., 2018, preprint (arXiv:1810.11394)
Cherenkov Telescope Array Consortium et al., 2019, Science with the Cherenkov Telescope Array. World Scientific Press, Singapore
del Valle M. V., Romero G. E., Luque-Escamilla P. L., Martí J., Ramón Sánchez-Sutil J., 2011, ApJ, 738, 115
Dermer C. D., Giebels B., 2016, Comptes Rendus Phys., 17, 594
Eislöffel J., 2000, A&A, 354, 236
Favata F., Flaccomio E., Reale F., Micela G., Sciortino S., Shang H., Stassun K. G., Feigelson E. D., 2005, ApJS, 160, 469
Feigelson E. D., Montmerle T., 1999, ARA&A, 37, 363
Flaherty K. M., Muzerolle J., 2008, AJ, 135, 966
Getman K. V., Feigelson E. D., 2021, ApJ, 916, 32
Getman K. V., Feigelson E. D., Garmire G. P., 2021, ApJ, 920, 154
Gibb A. G., 2008, in Reipurth B., ed., Handbook of Star Forming Regions. Vol. I: The Northern Sky. Astron. Soc. Pac., San Francisco, p. 693
Lada E. A., Depoy D. L., Evans Neal J. I., Gatley I., 1991, ApJ, 371, 171
Mazin D., et al., 2021, PoS, ICRC2021, 872
Nolan P. L. et al., 2012, ApJS, 199, 31
Reale F., 2003, Adv. Space Res., 32, 1057
Skinner S. L., Sokal K. R., Megeath S. T., Güdel M., Audard M., Flaherty K. M., Meyer M. R., Damineli A., 2009, ApJ, 701, 710
Spezzi L., Petr-Gotzens M. G., Alcalá J. M., Jørgensen J. K., Stanke T., Lombardi M., Alves J. F., 2015, A&A, 581, A140
Strom K. M., Strom S. E., Carrasco L., Vrba F. J., 1975, ApJ, 196, 489
Walther D. M., Geballe T. R., 2019, ApJ, 875, 153
Walther D. M., Robson E. I., Aspin C., Dent W. R. F., 1993, ApJ, 418, 310
Waterfall C. O. G., Browning P. K., Fuller G. A., Gordovskyy M., 2019, MNRAS, 483, 917
Wood M., Caputo R., Charles E., Di Mauro M., Magill J., Perkins J. S., Fermi-LAT Collaboration, 2017, PoS, ICRC2017, 824
Zhao B., Wang M., Yang J., Wang H., Deng L., Yan J., Chen J., 1999, AJ, 118, 1347

This paper has been typeset from a T$_{E}$X/L$^{A}$T$_{E}$X file prepared by the author.